\documentclass[aps,prl,twocolumn,superscriptaddress,nofootinbib,nobalancelastpage,showpacs,nobibnotes]{revtex4}
\usepackage{epsfig}

\begin{document}
\title
{Effects of ion and electron correlations on neutrino scattering in the infall phase of a supernova.
}

\author{R. F. Sawyer}
\affiliation{Department of Physics, University of California at
Santa Barbara, Santa Barbara, California 93106}

\begin{abstract}
Many authors have used one-component plasma simulations in discussing the role of ion-ion correlations in reducing neutrino opacities
during the collapse phase of a supernova.
In a multicomponent
plasma in which constituent ions have even a small range of N/Z ratios neutrino opacities are much larger,
in some regions of parameters, than for the case of a one component plasma. 

\pacs{95.30.Cq, 97.60.Bw}

\end{abstract}
\maketitle

There exists an extensive literature concerning the role of ion-ion correlations in reducing the
neutrino opacity in the region that collapses to form the supernova core \cite{i1}-\cite{janka}.
These correlations are important, by virtue of the fact that typical neutrino wavelengths are 
large compared to the Debye length.  Beginning with the
standard model result for the coherent neutrino scattering \underline{from a single ion} of charge Z and neutron
 number N, \footnote{We take $\hbar, c, k_B=1$ in what follows. }
\begin{eqnarray}
d\sigma_0 /d\Omega={G_W^2 C^2 E_\nu^2(1+\cos \theta) \over 4 \pi^2}\, ,
\end{eqnarray}
where
\begin{equation}
C=-2 Z \sin^2 \Theta _W+(Z-N)/2 \, ,
\end{equation}
it is standard to express the ionic correlation effects through a structure function $S(q)$. Then we write a 
differential scattering rate for the neutrino, $d \gamma / d\Omega$, as,
\begin{equation}
d \gamma(q)/d \Omega =(d \sigma_0 /d \Omega ) n_I \, S(q) \, ,
\label{rate}
\end{equation}
where $n_I$ is the ion number density and $\bf q$ is the momentum transfer to the ions.

The calculations of $S(q)$ reported in \cite{i1}-\cite{janka} are all based on a one component ionic plasma, the
electrons being sufficiently degenerate in the regions of interest as to form a virtually uniform background. In ref.\cite{janka}
the authors do attempt to simulate the multicomponent case by using effective averaged parameters in the one-component model.
But the present note provides evidence that in the regions in which $S(q)<<1$ and in which the nuclei have significant diversity 
in their $N/Z$ ratios, these one-component plasma simulations give neutrino scattering rates that are much smaller
than those that would be obtained in a genuine multicomponent calculation. 

We illustrate this for the case of a two component
plasma with average number densities $n_1$, $n_2$, and nuclear charges $Z_1$, $Z_2$. 
Taking into account the weak currents of the ions alone, we write the operative vector (coherent) part of the neutral current couplings to neutrinos, as, 
\begin{equation}
H_I=G_W\int d^3 x \, \psi_\nu^\dagger (x) \psi_\nu (x) [\lambda_1 Z_1 n_1(x) +\lambda_2 Z_2 n_2(x)] \, ,
\label{source}
\end{equation}
where 
\begin{eqnarray}
\lambda_{1,2}= {1 \over Z_{1,2}}[-2 Z_{1,2} \sin^2 \Theta _W+(Z_{1,2}-N_{1,2})/2]\, ,
\end{eqnarray}
and $n_1(x), n_2(x)$ are the respective density operators for the two varieties of ions.
We obtain,
\begin{eqnarray} 
d\gamma (q) /d\Omega={G_W^2  E_\nu^2(1+\cos \theta) \over 4 \pi^2} \times ~~~~~~~~~~~~~~~~
\nonumber\\
\int d^3 x\,  e^{i {\bf q \cdot  x}}\Bigr [Z_1^2\lambda_1^2 \langle  n_1(x) n_1(0)\rangle +
2 \lambda_1 \lambda_2 Z_1 Z_2 \langle n_1(x) n_2(0)\rangle
\nonumber\\
+\lambda_2^2 Z_2^2 \langle  n_2(x) n_2(0)\rangle \Bigr ]\, .~~~~~~~~~~~~~~~~~~~~~~~~~~
\nonumber\\
\,
\label{rate2}
\end{eqnarray}
The fact that $\lambda_1$ is different from
$\lambda_2$ for different nuclear species, even if only slightly different, is the key to what follows. 
The physical point  that
neutrinos scatter from the \underline{fluctuations} of the quantity on the RHS of
(\ref{source}), $[\lambda_1 Z_1 n_1(x) +\lambda_2 Z_2 n_2(x)]$. If $\lambda_1=\lambda_2$ then the fluctuations
of this source strength are proportional to the fluctuations in electric charge density, and the charge density does
not like to fluctuate in the $q=0$, or long range, limit. ( For the moment we assume that the high electron degeneracy prevents any fluctuation of the electron density.)  In the case of the two component plasma with $\lambda_1\ne \lambda_2$,
the source strength for neutrino scattering, which is \underline{not} proportional to the charge density,
 \underline{can} fluctuate while leaving the ionic charge
density strictly zero. 

To address this analytically, we begin in the $q \rightarrow 0$ limit, where a simple argument based on statistical mechanics suffices. Then
we shall give the solution of the two component problem for all $q$ at the Debye-H\"uckel (D-H) level
of approximation. For orientation, we begin with the statement from ref.\cite{i2} that
the ``correct small $q$ behavior" of the structure function $S$ is given for the one component
plasma by,

\begin{eqnarray}
S(q)=\Bigr [  {3 \Gamma \over (a_I q)^2} +{1\over k_B T} \Bigr ({\partial P \over \partial n} \Bigr )_T  \Bigr ]^{-1} ={q^2 \over \kappa^2}\,+\, O(q^4) \, ,
\label{correct}
\end{eqnarray}
where $\Gamma$ is the conventional plasma coupling constant and $a_I$ is the mean interionic spacing.
In the final equality we have substituted the expression for $\Gamma$ in terms the Debye wave 
number $\kappa^2=e^2 Z_I^2 n_I /T$.
The limit (\ref{correct}) illustrates the reluctance of the single component
plasma to fluctuate. 

For the two-component plasma we define the partial squared Debye wave numbers for the respective species 
as $\kappa_{1,2}=e^2 \,Z_{1,2}^2\, n_{1,2} /T$.
In this case (\ref{correct}) should be replaced by,
\begin{eqnarray}
\lim_{q\rightarrow 0} S(q^2)={T(\lambda_1-\lambda_2)^2 \kappa_1^2 \kappa_2^2  \over e^2  (\kappa_1^2+\kappa_2^2)n_I C^2}\, ,
\label{final}
\end{eqnarray}
where $C$ can now be taken as any average coupling constant factor for the two species. Note that the factor $ n_I C^2$
in the denominator cancels when we calculate the rate from (\ref{rate}). 

Eq.(\ref{final})
can be derived from
the basic principles of statistical mechanics, following the steps of section 2 , ``theory of multicomponent fluctuations", of
ref. \cite{rfs1} and enforcing in addition a
constraint of complete local neutrality of the plasma. 
We only sketch those considerations here, beginning from
the construction of the limit, in a purely classical treatment, of the mean
of quadratic forms in the Fourier components of fluctuations $\delta n_i(q)$,

\begin{eqnarray}
 \lim_{q\rightarrow 0} \langle \delta n_i (q) \delta n_j(-q) \rangle=
   T   {\partial n_i \over \partial \mu_j}\,(\rm Vol.)\, .
\label{fluct}
 \end{eqnarray}

For the free energy density functional needed to evaluate the derivatives with respect to the chemical 
potentials, $\mu_i$, we take just the kinetic term
plus the simplest term that ensures complete local neutrality when a parameter, $b\rightarrow \infty$. This
interaction is an energy per volume of $V=b (Z_1 n_1+Z_2 n_2-n_e)^2/2$.
The densities of species \#'s 1 and 2, are now given by,
\begin{eqnarray}
n_1=2\exp\Bigr [T^{-1}[\mu_1-b Z_1 (Z_1 n_1+Z_2 n_2-n_e)]\Bigr ]
\nonumber\\
\times \int{d^3 k\over (2\pi )^{3}}\exp (-k^2/2MT)\, ,
\label{oldstuff1}
\end{eqnarray} 
and
\begin{eqnarray}
n_2=2\exp\Bigr [T^{-1}[\mu_2-b Z_2 (Z_1 n_1+Z_2 n_2-n_e)]\Bigr ]
\nonumber\\
\times \int{d^3 k\over (2\pi )^{3}}\exp (-k^2/2MT)\,.
\label{oldstuff2}
\end{eqnarray}

After taking the logarithms of (\ref{oldstuff1}) and (\ref{oldstuff2}) we differentiate to obtain the matrix
$\partial \mu_i / \partial n_j$, which is then inverted to provide the RHS of (\ref{fluct}),
which when substituted in (\ref{rate2}) yields the result (\ref{final})in the $b\rightarrow
\infty$ limit.

To address the $q^2$ dependence at the D-H level we begin by defining a set of density correlators,
\begin{equation}
K_{i,j}(q)= T^{-1}\int d^3 x e^{ i {\bf q \cdot x}}\langle n_i({\bf x})n_j ({\bf 0})\rangle \,  ,
\label{correlator}
\end{equation}
as the polarization functions, where $T$ is the temperature.
In the absence of the Coulomb interactions of the particles in the plasma we would have simply $K_{i,j}=T^{-1}\delta_{i,j} \langle n_i \rangle$, that is, a diagonal matrix with number density values on the diagonal.

It is almost always useful to define ``proper" polarization parts; graphically in a perturbation development for the correlator these are sums of graphs for $K_{i,j}$ that are individually not divisible into
unconnected pieces by cutting one Coulomb line. 
The reconstitution of the complete correlator from these proper polarization parts, which we designate as $\Pi_{i,j}$, is fairly straightforward, and there are various languages for carrying it out. A complete and elegant derivation and statement 
of the results, in the case of a classical plasma, is given by Brown and Yaffe \cite{by}. We quote eq.(2.110) of this paper, with minor changes of notation, for the case of any number of ionic species,
\footnote{In the definition (\ref{correlator}) we added an extra factor of $T^{-1}$, as compared
to the definition in ref. \cite{by}. The development in ref. \cite{by} did not assume a uniform
neutralizing background, as we did; assuming instead a neutral system consisting of Boltzmann
particles of both signs of charge. However the formalism and equations are applicable to
our case, at the classical level, with the exceptions of sum rules based on neutrality
of the sea of particles that enter explicitly. Identical results can be obtained from the
usual finite temperature QFT approach to the many-body problem; we shall explicate the connection
in (\ref{RPA}) and the discussion that follows. In this latter approach the correlators $K(q,\omega)$ and their building
blocks $\Pi$ are the Fourier coefficients of $\tau$-ordered products ($\tau$=imaginary time), and the approximation
with the same structure as (\ref{general}) is called the RPA or the ring approximation.} 

\begin{eqnarray}
 e_a e_b K_{ab}(q)=e_a e_b \Pi_{ab}(q)-~~~~~~~~~~~~
\nonumber\\
\Bigr [\sum_c e_a e_c \Pi_{ac}(q) \Bigr ] \Bigr [\sum_c e_b e_c \Pi_{bc}(q) \Bigr ]G(q)\, ,~~~
\label{general}
\end{eqnarray}
where,
\begin{eqnarray}
G^{-1} (q)\equiv q^2+\sum_{a,b} e_a e_b \Pi_{ab}(q) \, ,
\label{g}
\end{eqnarray}
and $e_a\equiv e Z_a$ is the charge of nuclear species {\it a}. To obtain the Debye-H\"uckel level
result we insert the lowest order answer for the proper polarization parts and evaluate them at 
$q=0$, giving $Z_i^2 e^2\Pi_{i,j}(0)\rightarrow \delta_{i,j} \kappa_i^2$,
where $\kappa_i^2=e^2 Z_i^2 n_i/T$, the contribution of the $i$'th ionic species to the total Debye
screening (wave-number)$^2$. 
We should emphasize, however, that (\ref{general}) holds to all orders in the plasma coupling; a non-perturbative derivation 
of this result is given in ref.\cite{by}. 
\footnote{When we go beyond the lowest approximation, the off-diagonal parts of the
proper polarization matrix do not vanish.}

In the RPA approximation, we can, of course, employ simple graph-summing methods of quantum many-body
theory, as presented, for example, in ref. \cite{fw}, to regain the classical result (\ref{general}).
Going beyond the strictly classical, we can then add fluctuations in electron density to the picture;
the degenerate electrons were taken as a uniform gas in the derivation of (\ref{final}). At the DH level
we need only the square of the electron screening wave number,
\begin{equation}
\kappa_e^2=e^2 {\partial \over \partial \mu_e}n_e \approx e^2  \Bigr ( {3 \over \pi}\Bigr ) ^{1/3} n_e^{1/3}
\label{elec}
\end{equation}
where the last approximation is that of complete degeneracy.

Now we use (\ref{general}) for the case of the three species, two kinds of ions plus electrons, in the DH approximation, to obtain,
the ion-ion correlators,
\begin{eqnarray}
K_{1,1}={\kappa_1^2 \over e^2}\,\,{q^2+\kappa_2^2+\kappa_e^2  \over q^2 +\kappa_1^2+
\kappa_2^2+\kappa_e^2}\, ,
\nonumber\\
\,
\nonumber\\
K_{1,2}=-{ \kappa_1^2 \kappa_2^2 \over e^2}\,,~~~~~~~~~~~~~~~
\nonumber\\
\,
\nonumber\\
K_{2,2}={\kappa_1^2 \over e^2}{q^2+\kappa_1^2+\kappa_e^2  \over q^2 +\kappa_1^2+
\kappa_2^2+\kappa_e^2}\, .
\label{3spec}
\end{eqnarray}
We have not written down the expressions for the remaining three independent elements of the correlator matrix
$K_{e,e}, \,K_{1,e} \,K_{2,e}$, since we are still addressing only the effects of the electron fluctuations on
the baryonic fluctuations that determine the contribution of the baryonic current to the rates. If we were to
include the weak couplings of the neutrinos to electrons, there would be a piece of the amplitude
from neutrino-electron interactions that interferes coherently with the baryonic current, for very small values of
$q$, and we would need these correlators to calculate the effects. However, we shall argue below that
values of $q$ for which this occurs are too small for the interference to be of interest.

Using (\ref{3spec}) and (\ref{correlator}) in (\ref{rate2}) we obtain,

\begin{eqnarray}
d\gamma (q) /d\Omega=(4 \pi^2)^{-1}G_W^2  E_\nu^2(1+\cos \theta)W(q)
\label{answer1}
\end{eqnarray}
where,
\begin{eqnarray} 
W(q)=T \Bigr[ {\lambda_1^2 \kappa_1^2 (q^2+\kappa_e^2)
\over e^2 (q^2+\kappa_1^2+\kappa_2^2+ \kappa_e^2)}+~~~~~~~~~~~
\nonumber\\
\,
\nonumber\\
~~~~~~~~~{\lambda_2^2 \kappa_2^2 (q^2 +\kappa_e^2) \over e^2 
(q^2+\kappa_1^2+\kappa_2^2+ \kappa_e^2)}+{(\lambda_1-\lambda_2)^2 \kappa_1^2 \kappa_2^2 
 \over e^2 (q^2+\kappa_1^2+\kappa_2^2+\kappa_e^2)}\Bigr]\, .
\nonumber\\
\,
\label{answer2}
\end{eqnarray}

We note that in the free-particle limit of (\ref{answer2}), characterized by $q/\kappa \rightarrow \infty$, and for the case of ions with a common value of Z, we then obtain $S(q)=1$. But $q /\kappa$ remains quite small in any domain of interest in the
present problem. Note also that we recapture (\ref{final}) as the $q, \kappa_e\rightarrow 0$ limit of (\ref{answer2})

We estimate the numerical importance
of our of multicomponent effects for a case in which the density is $10^{12}$ g c$^{-3}$, the
composition, on the average, is nickel, and the temperature is $4$ MeV. For algebraic
simplicity we take the neutron number in both components to be the same and also set $\sin \theta_W=1/2$;
for the nuclear charges we take $Z=28 (1\pm \delta)$ so that $\lambda_{\pm}=1\pm \delta/2$. Then we have $\kappa_1^2 \approx 870\, (MeV)^2$ and
$\kappa_e^2 \approx 18 \, (MeV)^2$. The ratio of rate, with our effects included, to the one component plasma rate is then given to leading approximation in the parameters $\delta, \kappa_e$
by,
\begin{eqnarray}
{W(q,\delta, \kappa_e) \over W(q,0,0)}\approx 1+[1740 \,\delta^2 +18] \Bigr ({ 1{\rm MeV}\over q}\Bigr )^{2}.
\label{numbers}
\end{eqnarray}
Thus when the measure of nuclear diversity, $\delta$, is very small, we find a doubling of the 
one-component plasma result for a neutrino momentum transfer of $4$ MeV, coming from
the electron density fluctuations. For $\delta = .1$ we obtain a tripling. Larger diversity
parameters
will give much bigger enhancements, as would the choice of a smaller value of $q$. 

The point of this 
exercise was to evaluate the potential impact of our corrections, not to deal with a realistic mix 
of isotopes. In the real problem we have a complex mix of nuclei, and there are a wide
variety of possibilities for the components. In addition, in evaluating the potential impact,
we need to incorporate the whole scenario of neutrino production through electron capture 
in the dynamic environment just to know which regions of $q$ are most important. The part 
of the spectrum of neutrino momenta that dominates the lepton loss rate is dynamically determined, 
and it is clearly centered in a lower energy part of the spectrum than the $3\, T$ 
range that we think about in the usual energy transport problems. This comes about first through the fact that the opacity
is much less for the lower energy neutrinos, and secondly because of the
repopulation of these states through electron capture by nuclei.

Even worse, from the standpoint of using our results, is the fact that already in the parameter
region which we used in our numerical example above, the plasma is moderately strongly coupled, with 
a value $\Gamma=8.8$. In this case we expect the D-H results will become inaccurate as $q$ is increased 
beyond a certain point. Determining this point
requires a computational approach. Looking at the ``molecular dynamics" results of Luu et al \cite{luu}
for the one-component case, plotted in ref. \cite{burrows} for exactly the parameters which we
used above, we see, for example, that (for $T=4$ Mev)
D-H works fairly satisfactorily at $q=6$ MeV, but is low by a factor of ten at $q=18$ MeV.
Of course, since both the D-H and ``molecular dynamics" approaches will give very different
results in the multi-component case than in the single-component case, the guidance provided
by the above example could be regarded with suspicion. Thus we believe that an 
essential preliminary to doing real physics in this problem is to do Monte Carlo studies
of the correlators for a two component plasma, with both the diagonal and off-diagonal
terms in the density correlators determined with fairly high precision, since we now
understand that when we substitute these results into (\ref{rate2}) there will be near
cancelation of the terms, for small $q$. 

We return to the question of the role of electrons. As we saw in (\ref{numbers}),
the electron density fluctuations do lead to some appreciable effects in the case
of a single nuclear constituent, even in the absence of an electron-neutrino
coupling. The physics is that the densities of ions and electrons fluctuate together.
This comes without cost in Coulomb energy; this effect is inhibited by the
large bulk modulus for the degenerate electron sea, but still can be significant. 

There are also contributions
to opacity from electron-neutrino interactions. These have been estimated
in ref.\cite{hw}, and appear to be numerically small compared with the ionic
part, in the domain that we used for our above estimates. There are two reasons for
this: a) the ionic terms start with the advantage of the coherence factor,
of order $N$ (not $N^2$, since the electron number is larger than the
ionic number by approximately a factor of $N/2$); b) electron degeneracy 
severely limits the final electron states available in a neutrino scattering. That said,
we note that the calculations of ref.\cite{hw} which were carried
out for an electron gas at zero temperature, probably give a significant 
underestimate of the $\nu$-e scattering rate. If we are scattering thermal
neutrinos from a degenerate electron gas at temperature $T$, there are 
more final electron states available near the Fermi surface, due to the diffuseness of
this surface, than there would be from calculating the volume of the region
of non-overlap between a $T=0$ Fermi sphere and the sphere displaced by a momentum,
$\bf q $ where $q\approx T$. This becomes even more the case for sub-thermal
neutrinos. Thus improved calculations of the electronic current contributions
may be in order.

As we remarked before, we did not include any electron scattering contribution
in the coherent part of the calculation summarized in the result (\ref{answer2}).
Strictly speaking, it belongs there for very small momentum transfers,
$q <<T$, that is, when the energy transfer is negligible.
A more analytic form of this remark is that when the energy variable is introduced, the multi-component RPA equations
are still of the form (\ref{general}) where $\Pi_{i,j}(q,\omega)$ is now energy dependent.
However the rates are no longer given directly by integrals over $K_{i,j}(q,\omega)$
but rather by integrals involving,
\begin{equation}
{1 \over 1-\exp (-\omega/T)}\,{\rm Im} [K_{i,j}(q,\omega)]\,. 
\label{RPA}
\end{equation}

When the prefactor in (\ref{RPA}) can be approximated as $T/\omega$, then the energy
part of a phase space integration gives exactly the integral over the imaginary part
of the correlator which, through the dispersion relation, produces the real part evaluated at
zero energy, as in (\ref{rate2}). But for the values of $q$ that matter in the present case 
the expansion of the prefactor is not justified for the electron contributions, because a thermal
neutrino colliding with a relativistic electron will typically transfer an amount
of energy of order $T$. 

In any case, we find that in the regions of interest the ionic current
and electronic current contributions do not interfere very much, at least at the RPA level.
We add a caveat however; in a strongly coupled plasma we see no reason
for the rates coming from the two kinds of neutrino interaction to separate
so neatly. This could provide a further complication to a future Monte-Carlo
calculation aimed at settling the issues raised in the present paper.

There is a close relation between the above developments on neutrino scattering and some important corrections to Compton scattering in a hydrogen plasma. Indeed, the photon-electron interaction that produces the Thomson limit
is almost identical in form to the neutrino-ion interaction in the present paper.  The calculation again demands the careful consideration of a two component plasma, and the mechanics
is parallel to that presented in the present paper. The effects are actually important in the calculation of Compton opacities in the solar interior. This subject was discussed
in a number of references over the years. Boercker \cite{boercker} carried out the calculation that appears to be completely correct, obtaining significant corrections that are incorporated into the solar opacity codes that are in use today. When we go to slightly more extreme conditions than those in the solar
interior, a density of 1 gc$^{-3}$ and temperature of $10^6$K, for example, in a hydrogen plasma,
we find, using the analog of (\ref{final}), 
\begin{equation}
S(q_{\rm therm})\approx S(0)={\kappa_I^2 \over (\kappa_I^2+\kappa_e^2)}={1\over 2}\, ,
\label{sun}
\end{equation}
where now $\kappa_I \approx \kappa_e$.\footnote{In the solar interior we are not in the region in which the limiting form $q\rightarrow 0$
can be applied. Ref. \cite{boercker} deals with the complete $q$ dependence.}

Recapitulating some of the conclusions of this paper, we found that the $q=0$ limit of the
structure function is generally non-zero, in contrast to the conclusions of the
large literature that uses ``effective" one-component plasmas of some kind. The
answer for $S(0)$, for the case of frozen electrons and any number of ionic
species, can be found simply from energetic arguments using basic statistical mechanics,
and even when it is relatively small it protects against the dramatic suppressions
found in the current literature. We note that the results are in complete contradiction to the results of
the procedures for ionic mixtures proposed in ref. \cite{i2} and recently used in ref. \cite{janka}.
For application to supernovae, we need the extension to finite $q$ in the strong coupling regime.
This should begin with the numerical investigation of the reliability of the RPA results (\ref{answer2})
for the case of a classical plasma with two ionic components and for a variety of plasma coupling
strengths.


\begin{thebibliography}{00}
\bibitem{i1} N.Itoh, Prog. Theor.Phys. {\bf 54}, 1580 (1975)
\bibitem{bw} R Bowers and J. R. Wilson, Astrophys. J {\bf 263}, 366 (1982)
\bibitem{i2} N. Itoh,{\it et al.}, Astrophys. J. {\bf 611 }, 1041 (2004)
\bibitem{bm} S. W. Bruenn and A. Mezzacappa, Phys. Rev.{\bf D56}, 7529 (1997)
\bibitem{h1} C. J. Horowitz, Phys. Rev. {\bf D55}, 4577 (1997)
\bibitem{lieb} M. Liebendoerfer, M. Rampp, H.-Th. Janka, A. Mezzacappa, Astrophys. J {\bf 620}, 840 (2005)
\bibitem{janka} A. Marek , H.-Th. Janka , R. Buras , M. Liebendoerfer , M. Rampp, astro-ph/0504291
\bibitem{rfs1} R. F. Sawyer, Phys. Rev. {\bf C40}, 865 (1989)
\bibitem{by}L. S. Brown and L. G. Yaffe, Phys. Repts. {\bf 340} nos. 1-2  (2001)
\bibitem{fw}A. Fetter and J. D. Walecka, {\it Quantum Theory of Many-Particle Systems} (McGraw-Hill, New York, 1971)
\bibitem{luu} T. Luu, A. Hungerford, J. Carlson, C. Fryer, and S. Reddy (to be published)
\bibitem{burrows} A. Burrows, S. Reddy, T. Thompson, astro-ph/0404432
\bibitem{hw} C. J. Horowitz and K. Wehrberger, Phys. Rev. Lett. {\bf 66},272 (1991)
\bibitem{boercker}  D. B. Boercker,  Astrophys. J. {\bf  316}, L95  (1987)
\end{thebibliography}
\end{document}